\documentclass[12pt,preprint]{aastex}

\newcommand{\myemail}{quanz@mpia.de}

\slugcomment{To appear in ApJ.}

\shorttitle{Dust rings and filaments around V1331 Cygni}
\shortauthors{Quanz, Apai \& Henning}

\begin{document}



\title{Dust rings and filaments around the isolated young star V1331 Cygni\footnote{Based on observations made with the NASA/ESA Hubble Space Telescope, obtained from the data archive at the Space Telescope Institute. STScI is operated by the association of Universities for Research in Astronomy, Inc. under the NASA contract  NAS 5-26555.}}


\author{S. P. Quanz\altaffilmark{1}}
\affil{Max Planck Institute for Astronomy, K\"onigstuhl 17, 69117 Heidelberg, Germany}
\email{\myemail}

\author{D. Apai}
\affil{Steward Observatory, 933 North Cherry Avenue, Tucson, AZ 85721, USA\\
NASA Astrobiology Institute, Mail Stop 240-1, Ames Research Center, Moffett Field, CA 94035, USA}

\and

\author{Th. Henning}
\affil{Max Planck Institute for Astronomy, K\"onigstuhl 17, 69117 Heidelberg, Germany}




\begin{abstract}
We characterize the small-- and large--scale environment of the 
young star V1331 Cygni with high--resolution \emph{HST/WFPC2} and Digitized Sky Survey images. 
In addition to a previously known outer dust ring ($\approx$30$''$ in diameter), the \emph{HST/WFPC2} scattered light image reveals an inner dust ring 
for the first time. This ring has a maximum radius of $\approx$6.5$''$ and is possibly related to a molecular envelope. 
Large-scale optical images show that V1331 Cyg is located at the tip of a long dust filament linking it to the dark cloud LDN 981.
We discuss the origin of the observed dust morphology  
and analyze the object's relation to its parent dark cloud LDN 981. 
Finally, based on recent results from the literature, we investigate the properties of V1331 Cyg 
and conclude that in its current state the object does not show
sufficient evidence to be characterized as an FU Ori object.

\end{abstract}





\section{Introduction}
Stars form from collapsing molecular cloud cores possibly triggered by different mechanisms:
turbulence 
may lead to overdensities in molecular cloud cores, or external triggers, such as supernovae shock waves or nearby 
violent stellar winds, may initiate the collapse of such cores. 
After the initial collapse and the formation of a central object 
low-- and intermediate--mass stars (several tenths to a few solar masses) 
continue to accrete matter via a circumstellar disk. The disk is
replenished with matter from a remaining envelope in the early stages. 
This phase might coincide with the FU-Orionis (FUOR) phenomenon observed for a small group of young, low--mass objects 
\citep[for a review see,][]{hartmann1996}.


In this paper we focus on the young star V1331 Cyg. Due to the similarity of its spectrum
to that of the well--known FUOR V1057 Cyg prior to its outburst, V1331 Cyg was classified
as a pre--outburst FUOR candidate by \citet{welin1976} and also \citet{herbig1989}. 
Further support for this classification came from the ring--like reflection nebula surrounding V1331 Cyg \citep{kuhi1964} that 
appeared very similar to those found around other FUORs \citep{goodrich1987}.
Due to a lack of photospheric absorption lines 
the spectral type of V1331 Cyg was mostly derived from photometric measurements and 
covers a range from G0 \citep{kolotilov1983}, through F0-F2 \citep{mundt1981} and F0-A8 \citep{chavarria1981} to B0.5 \citep{cohen1979}.  
Also the distance towards V1331 Cyg is uncertain and literature estimates range from 694 pc  \citep{chavarria1981} to 550 pc \citep{shevchenko1991}. 
We adopt the conservative distance of 550 pc which is in agreement with most recent publications 
\citep[e.g.,][]{mcmuldroch1993,mundt1998,henning1998}.
Strong P Cygni absorption in many of its lines suggested the presence of circumstellar matter surrounding the object \citep{chavarria1981,mundt1984}. A compact circumstellar disk was suggested by Weintraub et al. (1991) based on submillimeter continuum observations. 
\citet{mcmuldroch1993} found additional evidence for the existence of a massive 0.5 M$_\sun$ disk 
surrounded by a flattened gaseous envelope from CO synthesis maps. Furthermore, these authors identified 
a bipolar outflow and a radially expanding gaseous ring containing more than 0.07 M$_\sun$. 
They suggested that this ring was a swept-up gaseous torus from an energetic mass ejection stage that was probably
linked to one or several FUOR outburst(s) $\sim 4 \times 10^3$ years ago.

In the following we present and discuss high--resolution images of the 
dusty circumstellar environment of V1331 Cyg. 
The possible origin of the observed dust structure, the object's relation to 
its parent molecular cloud and possible external triggers for the star formation process are discussed. 
We also investigate the FUOR-like nature of V1331 Cyg. 






\section{Data reduction}
V1331 Cyg (RA: $21^h 01^m 09.21^s$, DEC: +50$^{\circ}$ 21$'$ 44.8$''$, J2000) was observed with \emph{HST/WFPC2} in August 2000. 
Two exposures with 230 seconds each were taken in the F606W filter 
with the WF3 camera. We performed simple image processing and cosmic ray removal. 
The center of the star was saturated leading to overflowing columns along the y-axis of the detector  (roughly NNE to SSW in Fig.~\ref{figure1}).
A simulated \emph{WFPC2} Point Spread Function (PSF) was created using the TinyTim software package \citep[][version 6.3]{krist1993}\footnote{http://www.stsci.edu/software/tinytim/}.
This PSF was subtracted from the original image in order to reveal faint circumstellar structures.
The accurate scaling of the reference PSF was not straightforward due to the saturation of the original image. 
We minimized the residuals in the center of the star and along the
diffraction spikes, which simultaneously allowed an accurate positioning of the PSF.
For analyzing the large scale environment we additionally downloaded \emph{Digitized Sky Survey 2 (DSS2)} data from the ESO Online Digitized Sky Survey homepage\footnote{http://archive.eso.org/dss/dss}.

\section{Results}

\subsection{Small--scale morphology (HST/WFPC2)}
\begin{figure}
\centering
\plotone{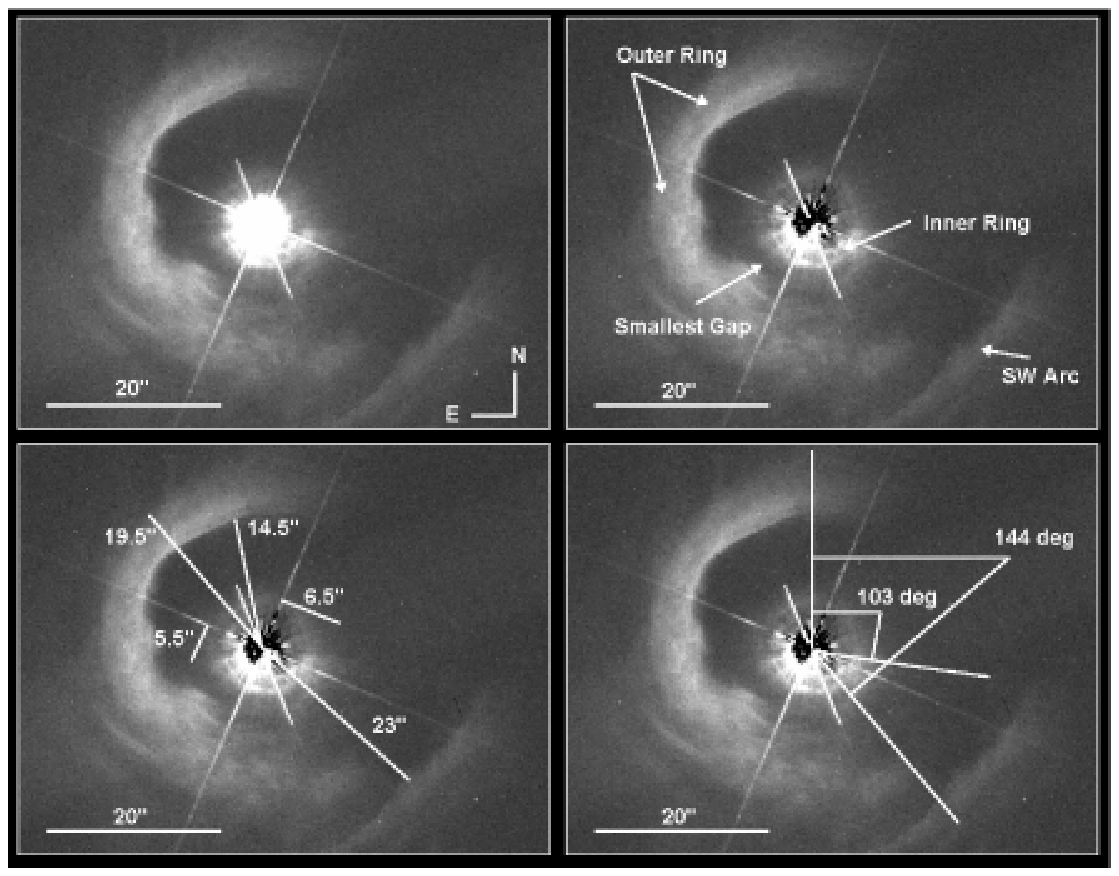}
\caption{Cleaned original image (upper left) and PSF-subtracted image (upper right) of V1331 Cyg 
in the WFPC2 F606W filter. The dusty environment
is resolved and the outer and inner dust ring are clearly visible. Selected radii 
of the dust rings and opening angles for
the dust free regions are shown in the lower left and lower right panel, respectively 
(north is up, east to the left in all images). \label{figure1}
}
\end{figure}

Fig.~\ref{figure1} shows the resulting PSF-subtracted image for the F606W exposures. The high resolution of the \emph{WF3} camera (1.22\,$\lambda/D\approx0.064''$ with a pixel scale of $\sim 0.1''$/pixel) reveals the structure of the dusty circumstellar environment 
in great detail: In addition to the large dust ring already mentioned by \citet{kuhi1964} we find for the first time direct evidence for an additional ring-like dust structure close to the central star. In the following we refer to the former ring as the "outer ring", call the latter one "inner ring" and refer to a faint arc in the south-west (roughly 23$''$ from the central star) as the "SW-arc".

For the outer ring we computed a mean surface brightness of 
21.12 mag/arcsec$^2$ in the Vega magnitude system \citep{gonzaga2002}. 
The inner edge of its north-eastern part is well--defined, while it becomes more fuzzy as the ring continues southwards (Fig.~\ref{figure1}). 
In the north-east the distance between the inner edge of the outer dust ring and the central star is $\approx$14.5$''$, while
the outer radius of the ring is $\approx$19.5$''$. 
From north--west to south--west the
outer ring is intersected over an angular range of $\approx$144$^{\circ}$ (not taking the SW-arc into account) and no scattered light is visible. 

The newly resolved inner ring is separated from the outer ring by a well--defined gap with a 
smallest projected distance of  $\approx$2.6$''$ south-east of the central object.
The inner dust ring extends almost exactly half a circle around V1331 Cyg from the north-east to the south-west 
and scattered light can be traced very close to the center. 
The inner ring is mostly visible in regions where the inner edge of the outer ring 
appears rather fuzzy. 
The outer radius of the inner ring is $\approx$5.5$''$ measured along the diffraction spike in south-east direction and $\approx$6.5$''$ measured along the diffraction spike in south-west direction (lower left panel of Fig.~\ref{figure1}).

Assuming a distance of 550 pc the radius of the inner dust ring corresponds to
$\approx$3,000 AU and $\approx$3,600 AU in the south-east and south-west direction, respectively,  
the inner radius of the outer dust ring is roughly 8,000 AU, and the
distance between the central object and the faint SW-arc is $\approx$13,000 AU.


\subsection{Large--scale morphology (DSS)}
\begin{figure}
\centering
\epsscale{1.0}
\plotone{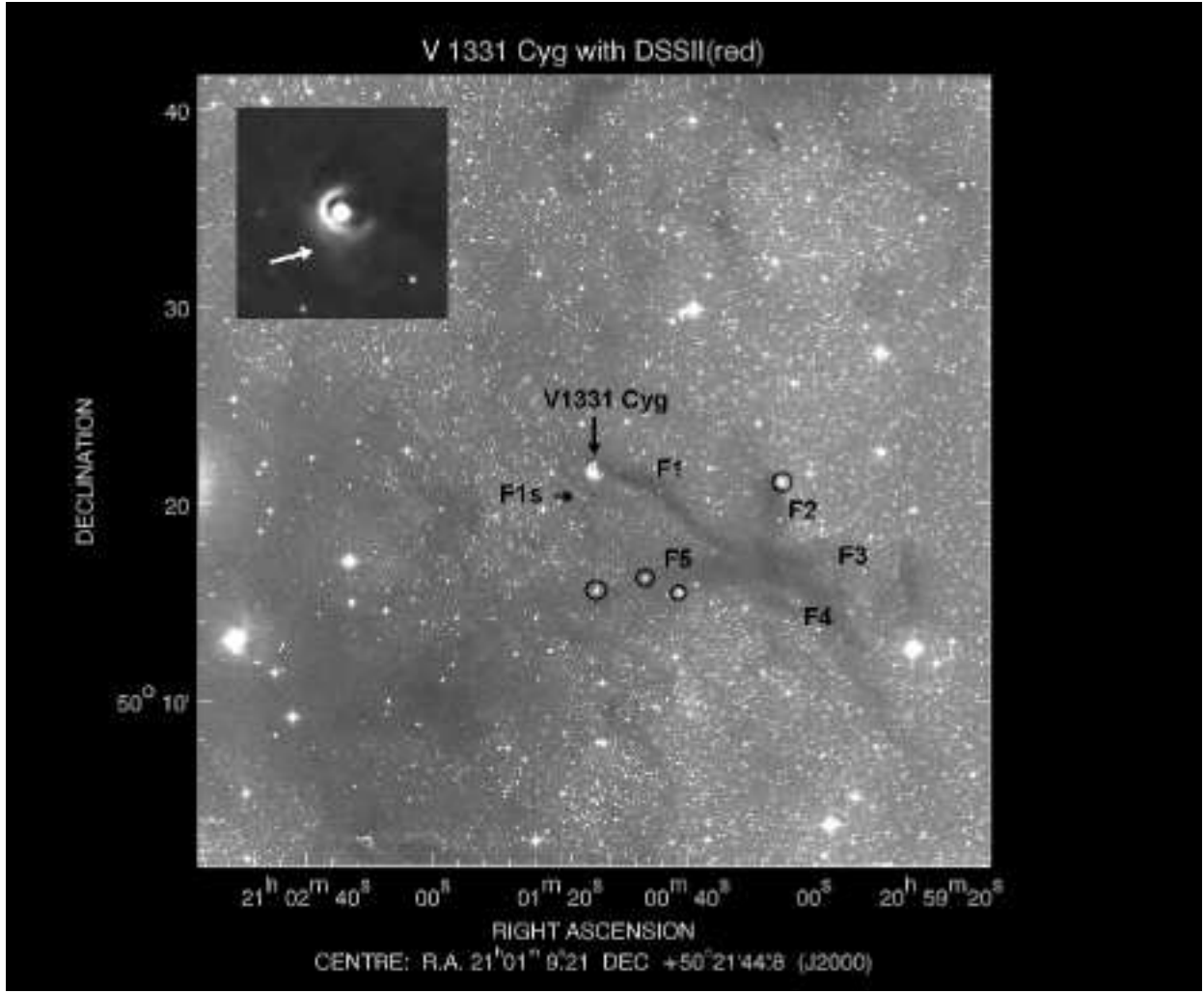}
\caption{Red image 40'x40' in size from the DSS2 data archive. The position of V1331 Cyg is indicated by the arrow. 
The bright outer ring of V1331 Cyg is clearly visible and also its apparent 
connection to a dark filament (F1) stretching in and intersecting
the dust ring in the north west. This filament is part of the "hand-like" dark cloud LDN 981 which has
its center south west of V1331 Cyg. 
Also visible is an additional fainter and more wiggled filament streching southwards
from V1331 Cyg. A weak signal of scattered light (indicated by the arrow) can be seen along the edge of this filament in 
the close-up inset image taken with the blue filter of DSS2. Finally, we indicated some stars from
the H$\alpha$ survey by \citet{feigelson1983}. These objects were not classified as pre-main-sequence stars, although
their position relative LDN 981 might suggest that they recently evolved from the filaments. \label{figure2} }
\end{figure}

Fig.~\ref{figure2} shows an R--band image from the \emph{DSS2} data archive.
The image is 40$'\times$40$'$ in size and centered on V1331 Cyg. The object is located at the eastern tip of a $\approx$8$'$ long 
dark filament (F1 in Fig.~\ref{figure2}). The filament links V1331 Cyg to the dark cloud Lynds 981 \citep[LDN 981, RA: $20^h 00^m 15^s$, DEC: +50$^{\circ}$ 17.3$'$, J2000;][]{lynds1962}. The cloud consists of a roughly elliptical core with five elongated dark filaments stretching radially outwards to which in the following
we refer to as F1--F5. 
The outer dust ring of V1331 Cyg is clearly visible in the \emph{DSS2} 
image and the dark filament F1 seems to intersect the outer dust ring in the north--west. Furthermore, a fainter filament (F1s) seems to run southwards from the object.


\section{Discussion}
\subsection{Possible origin for the dust rings}\label{origin_of_rings}
The direct vicinity of V1331 Cyg is certainly shaped by the star and hence provides information 
on recent stellar activity. The observed morphology is defined by a prominent, symmetrical pair of rings seen in scattered light.
To explain the apparent morphology of the previously known outer dust ring it was suggested that  
we are looking into a cone shaped wind--blown cavity with its axis parallel 
to the star's (and disk's) rotation axis \citep{mundt1998}. The unresolved circumstellar disk
should thus have an almost pole-on orientation.
This picture is supported by the observations of a
bipolar molecular outflow whose relative strength implies that V1331 Cyg is viewed nearly 
pole-on \citep{mcmuldroch1993}. Furthermore, from observed P-Cygni profiles in H$\alpha$ and Na I \citet{mundt1984} and \citet{mundt1998} derived corresponding wind velocities that showed that the
outflow is oriented relatively close to the line of sight.

The existence of strong winds creating the observed cavity was also proposed to explain an expanding torus of molecular gas \citep{mcmuldroch1993}.
This torus lies outside the outer dust ring following its outer edge closely from the north--east to the south--west. 
\cite{mcmuldroch1993} computed a momentum between $1.5\;M_\sun kms^{-1}$ and $11.9\;M_\sun kms^{-1}$
for the gas torus and concluded that the current mass loss rate ($\dot{M} \le 10^{-6}M_\sun yr^{-1}$)
and wind velocities ($\approx 370\;kms^{-1}$) \citep{mundt1984} are not sufficient to sweep up the torus within its 
assumed lifetime of $\sim 4\times 10^3 yr$. They speculate that the torus was created in a more vigorous stage of V1331 Cyg, possibly during an FUOR event.
Since such an outburst is attributed to an increase in the mass accretion rate of the 
circumstellar disk which in turn leads to an 
increase of the system's mass loss rate (e.g., due to disk winds\footnote{Typical values are 
$\dot{M}_{wind}/\dot{M}_{acc}=0.1$ \citep[e.g.,][]{koenigl1998}}) such an event might explain the ejection of the torus and the creation of the well--shaped wind--blown cavity we observe in our scattered light image today.



\begin{figure}
\centering
\includegraphics[scale=1.0]{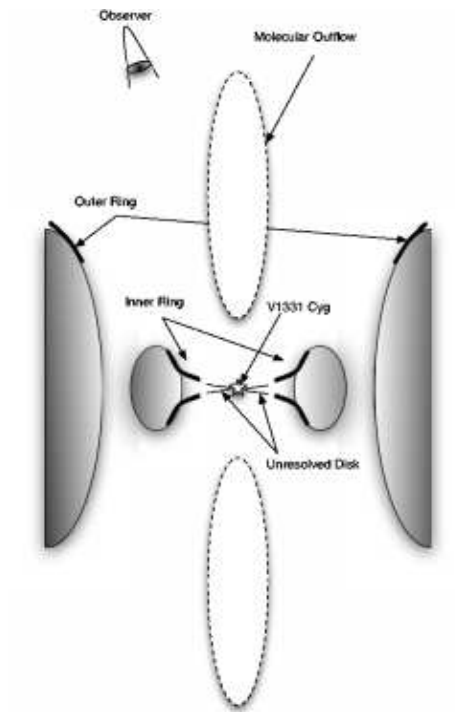}
\caption{Simplified sketch of the immediate surroundings of V1331 Cyg explaining the observed ring-like structures.\label{sketch}
}
\end{figure}

For the inner dust ring we believe that the scattered light \emph{HST} image reveals the 
dusty counterpart to a molecular structure which is surrounding the inner star-disk system.
The dimensions of the newly discovered inner ring are very similar to those of a flattened CO envelope 
found by \citet{mcmuldroch1993}. The assumed circumstellar disk seems to be more compact (source size less than 2000 AU) as it was unresolved in 1.3 mm dust continuum observations \citep{mcmuldroch1993,henning1998}.
Fig.~\ref{sketch} shows a simple sketch of the circumstellar environment of V1331 Cyg that is in agreement with most observational data.



\subsection{V1331 Cyg and LDN 981}
Fig.~\ref{figure2} shows that the dark filament LDN 981-F1 is responsible for the observed 
break in the outer dust ring. Either the dust filament
lies in the foreground and by chance in the line--of--sight toward V1331 Cyg, or 
V1331 Cyg itself recently emerged from the 
filament.

The most direct evidence for the filament being directly associated with V1331 Cyg is the fact that V1331 Cyg illuminates 
part of the filament running southwards (see, Fig.~\ref{figure2} inset). 
The direct association is further supported by the identical line--of--sight velocity of the LDN complex 
and the envelope of V1331 Cyg: \citet{park2004} measured the $^{13}$CO
velocity component of LDN 981-1 and LDN 981-2 (RA: $21^h 00^m 13.2^s$, DEC: +50$^{\circ}$ 20$'$ 06.0$''$ and 
RA: $21^h 00^m 15.0^s$, DEC: +50$^{\circ}$ 16$'$ 56.0$''$, J2000) to be $v_{LSR}=0.1\;kms^{-1}$
and $v_{LSR}=0.5\;kms^{-1}$ with $\Delta v=1.12\;kms^{-1}$ and $\Delta v=1.73\;kms^{-1}$,
respectively. \citet{levreault1988} derived a CO velocity in the direction of 
V1331 Cyg of $v_{LSR}=-0.7\;kms^{-1}$.
Based on these observations V1331 Cyg has to lie close to the edge of the two filaments (F1 and F1s), 
most likely at their intersection.
If so, V1331 Cyg has probably evolved from the filaments and is in the process of dispersing the filaments through its winds.


\subsection{Isolated or triggered star formation?}
Although V1331 Cyg is well established as a young star and the relation to 
LDN 981 is convincing it seems as if no additional star formation is going on within LDN 981 \citep{alves1998}.
In an earlier H$\alpha$ survey \citet{feigelson1983} found no evidence for additional pre--main--sequence stars  in an area of 20$'\times$20$'$ ($\approx$3.2$\times$3.2 pc) centered on V1331 Cyg. In Fig.~\ref{figure2} we marked some objects from that survey that appear to be related to LDN 981. However, they did not show strong
H$\alpha$ emission and were thus not classified as pre--main--sequence stars.
According to the VizieR database \citep{ochsenbein2000} these objects were also not detected 
in UV surveys and do not show a very prominent NIR-excess emission in the 2MASS database which would have been indicative of a young age.
 
Apart from having no additional young stars nearby, molecular line observations suggest that the parent cloud LDN 981 itself is stable and without on--going star formation, too: From CS(3-2) and DCO spectral line profiles \citet{lee2004} found no evidence for
infall signatures. However, as these authors pointed only towards the center of the filament LDN 981-F2 (see, Fig.~\ref{figure2}, RA: $21^h 00^m 13.1^s$, DEC: +50$^{\circ}$ 20$'$ 05.9$''$, J2000) with a beam of 43$''$ (FWHM) these results might not be applicable to the whole dark cloud (see also section 4.4).

The region closest to LDN 981 where evidence for recent star formation is found is LDN 988. Recently, \citet{herbig2006} analyzed the young stellar cluster associated with the Be star LkH$\alpha$324 (RA: $21^h 03^m 54.2^s$, DEC: +50$^{\circ}$ 15$'$ 09$''$, J2000) for which they assume a distance of 600 pc
and which lies east of V1331 Cyg at a projected distance of $\approx 30'$ (4.8 pc). Although the stars do not form a group following a single well-defined pre-main-sequence model, theoretical isochrones suggest a median age of 0.8 Myr for the population. 

It is worth to investigate, whether there is any evidence for a casual link between the star formation process in LDN 981 and LDN 988. Especially, since there is data suggesting that LDN 981 might be in a gravitationally stable configuration, it is not clear why and how V1331 Cyg should have formed at all.
In this respect we identify and analyze three possibilities: First, a propagating density wave 
triggered the star formation in LDN 988 and continued to travel in the direction of V1331 Cyg; second, a nearby 
supernova (SN) explosion triggered simultaneously the formation of V1331 Cyg and the stellar cluster in LDN 988 via shock waves; 
and third, a high proper motion early--type O-- or B--star passed by and its winds initiated the star formation 
process. The possibility of spontaneous star formation in the filaments is discussed in section 4.4. 

In the first case, assuming a local speed of sound of $0.3\;kms^{-1}$ within the 
molecular cloud, any subsonic pressure wave would need at least 15 Myr to reach from LDN 988 to LDN 981, 
an order of magnitude longer than the estimated age of the cluster. Thus, 
shock waves would need to travel roughly two orders of magnitudes faster than the value above in 
order to create a casual link between the two regions and this seems highly unlikely. 

To identify a possible SN explosion as an external star-formation trigger, we searched the available SN-remnant catalogues \citep[e.g., catalogue version 2001 of][]{green1996} in a radius of 3$^{\circ}$ around V1331 Cyg. We found only one candidate, G089.0+04.7 (HB21), with a projected distance of $\sim$2.58$^{\circ}$. However, \citet{byun2006} estimated the distance of G089.0+04.7 to be 1.7 kpc and derived
an age for the remnant of 1.5$\times$10$^4$ yr scaled to this distance. This supernova remnant is hence too young and also too distant to have triggered the formation of V1331 Cyg. 

Finally, we searched for early--type O-- and B--stars in the vicinity of V1331 Cyg
that could have triggered its formation via stellar winds. Based on the Hipparcos proper motion catalogue \citep{perryman1997}, we identified 6 objects within a radius of 3$^{\circ}$ with spectral types between B1 and B8. We plotted the projected trajectories the six stars have travelled during the last 1 Myr and found only one candidate (HIC 103530) that passed by V1331 Cyg as close as $\approx$0.4$^{\circ}$ ($\approx$3.8 pc at the distance of V1331 Cyg). From the Hipparcos parallax (2.58$\pm$0.65 mas) we derived a maximum distance to HIC 103530 of $\approx$518 pc 
which is at the lower limit of the distance estimates for V1331 Cyg. 
However, given the spectral type of HIC 103530 (B5Vn) and its minimum separation of  3.8 pc from its 
position relative to V1331 Cyg it seems unlikely that the star was triggering the formation of V1331 Cyg.

In summary, all observations suggest that V1331 Cyg formed in  isolation. 


\subsection{Star formation in filaments}
Large dust filaments are common structures in star--forming regions. Within these filaments sometimes denser fragments can be found possibly arising from collapsing globules that eventually will form new stars \cite[see, e.g.][]{schneider1979, apai2005}.
Analyzing the spatial distribution of young stars in the Taurus molecular cloud
\citet{hartmann2002} found that 
most of these young objects are aligned along three filaments following the distribution of denser gas and dust. 
The mean separation between the stars along the filaments was found to be about 0.25 pc,
consistent with a roughly uniform density along the filaments. This value also compares well to the estimated Jeans length $\lambda_J$ in the densest gaseous filaments\footnote{Typically, one finds $\lambda_J = 0.19\,pc\,(\frac{T}{10\,K})^{1/2}(\frac{n_{H_2}}{10^4cm^{-3}})^{1/2}$.} \citep{hartmann2002}. 

A direct comparison between these examples and the dark cloud LDN 981 is difficult. We lack appropriate data
to determine whether we find any denser globules within the dark filaments, and also the spatial 
extension of the filaments, e.g.,  in Taurus, is much larger (partly more than 12 pc) and numerous stars have already formed. 
However, we note two important points: First, based on the estimates for the typical Jeans length $\lambda_J$
in Taurus, one might expect additional stars to form also within the filaments of LDN 981. And second, while the 
three major filaments in Taurus lie more or less parallel to each other with young stars almost spaced equally 
along the filaments, LDN 981 shows filaments reaching radially outwards from an elliptical center with one young star
at the tip of one filament. This difference in the morphological structure of the observed filaments might suggest a different origin for the filaments. Recently, \citet{burkert2004} presented 2D simulations of collapsing, 
self-gravitating gaseous "sheets" with different initial geometries. Collapsing elliptical sheets tend to produce filaments with major mass concentrations at the ends, but with additional sub--fragments spaced along the filament (similar to Taurus). 
The filamentary structure of LDN 981, on the other hand, has resemblance to results from simulations of a collapsing "ghost"-sheet with highly irregular boundaries \citep[see, Fig. 12 of][]{burkert2004}. In this case, stars form first at the ends of the filaments and, in addition, the first stars are thought to form at the
end of the thinnest and longest filaments, which is exactly what we observe in LDN 981. 
Thus, even if this model is still very limited and simple it might explain the differences in the observed filamentary structures and also the formation of V1331 Cyg at the end of the very elongated filament LDN 981-F1. One problem arises, though: \citet{burkert2004} modeled \emph{collapsing} gaseous sheets and there seem to be indications that infall is not present in LDN 981 \citep{lee2004}.
However, this result is based on a single 
pointing with moderate spatial resolution towards the center of LDN 981-F2 which might not be representative for the 
whole cloud (see, also section 4.3). The theoretical model predicts stronger infall signatures along the edges of the filaments as well as along the thinner and longer filaments. A more sensitive and spatially larger search for these signatures within the filaments in LDN 981 may hence be worthwhile. 

It seems as if LDN 981's picture as a collapsing cloud core might be able to explain both the filamentary structure and the formation of V1331 Cyg at the tip of the most elongated filament. To test this hypothesis, future studies of the whole dark cloud should address the issue of additional denser globules within the filaments of LDN 981 and further (sub-)millimeter observations can be used to better constrain the dynamics of the filaments.

\subsection{V1331 Cyg - an FU Ori object?} 
\label{sectionFUOR}
V1331 Cygni was among a group of ten young stars initially defined as pre--outburst FUORs by \citet{welin1976}. 
And as described in section~\ref{origin_of_rings}, the circumstellar environment of V1331 Cyg
bears indeed evidence of a more active evolutionary phase some thousand years ago, which possibly 
puts V1331 Cyg in between two consecutive outburts.
Based on recent observational results
from the literature we shortly investigate whether V1331 Cyg still fulfills the criteria of an 
FUOR even in its \emph{present} stage.

FUORs share in general several common features summarized in \citet{hartmann1996}. For the first members of the
group a rapid outburst in optical and NIR luminosity of several magnitudes was observed. The outburst is
followed by a decrease in luminosity which can last between serveral tens to hundreds of years.
However, the photometric behavior of V1331 Cyg over the last 40 years 
shows no evidence for a significant increase or decrease in NIR flux \citep{abraham2004}. 

Apart from the outburst which might be difficult to observe, the best 
criterion to identify an FUOR is its spectrocopic behavior. While modest resolution 
optical spectra reveal spectral types of late F to G supergiants most
infrared features are best fit with K-M giant-supergiant atmospheres. 
Although no clear spectral type could thus far be derived for V1331 Cyg most
authors agree on a spectral type earlier than F2 \citep{cohen1979,mundt1981,chavarria1981}. 
Only one author found a spectral type of G0 \citep{kolotilov1983}.

Furthermore FUORs show strong CO bandhead absorption at 2.3$\mu$m while
V1331 Cyg shows unambiguously varying CO bandhead emission 
\citep{biscaya1997}.

Thus, V1331 Cyg can currently not be classified as an FUOR object and seems to be in 
a more quiescent evolutionary phase.

\section{Summary and Conclusions}
We presented high--resolution scattered light images of the dusty environment of V1331 Cyg.
Two circumstellar dust rings, well separated by a gap, are detected
with the inner ring being resolved for the first time. This dust ring is 
probably related to a known flattened CO structure several thousand AU in size.
The outer ring--like dust structure likely represents the
remnants of the molecular core V1331 Cyg formed from. As V1331 Cyg is
surrounded by a circumstellar disk seen close to pole--on stellar and/or disk winds and
outflows might have created a cavity towards the line of sight of the observer.
The large scale structure suggests that V1331 Cyg is physically 
related to the dark cloud LDN 981 and lies at the edge in between two 
dark filaments where it recently emerged from. The distance estimates for V1331 Cyg and LDN 981 
are in good agreement.
V1331 Cyg is likely to be the only young star in LDN 981 and did presumably form
in isolation.
Based on recent results of 2D simulations of collapsing irregular gaseous sheets that appear very similar to 
the morphology we observe for LDN 981, we suggest that LDN 981 is undergoing a gravitational collapse and 
may form additional stars along the filaments in the future.
We did not find any evidence for a possible external trigger for the formation of
V1331 Cyg, such as propagating shock waves from a nearby supernova remnant or 
high proper--motion early--type stars in the vicinity.
Finally, observational results from the last decades show that V1331 Cyg does not
fulfill the criteria for an FUOR in its current status. However, more vigorous 
events such as FUOR outbursts might have occured in the past and helped
creating the interesting dust morpholgy we observe today.




\acknowledgments
Sascha P. Quanz kindly acknowledges support from the German \emph{Friedrich-Ebert-Stiftung}. 
Part of this work was supported by NASA through the NASA Astrobiology Institute under
Cooperative Agreement No. CAN-02-OSS-02. We thank M. Andersen, S. J. Kim, I. Pascucci, J. Muzerolle,
O. Krause, H. Linz, H. Beuther and R. Mundt for helpful discussions.
This research has made use of the SIMBAD database,
operated at CDS, Strasbourg, France.

\clearpage

\clearpage

\clearpage

\end{document}